%% file: moriond-chaty.tex
\documentclass[11pt]{article}
\usepackage{moriond,epsfig}
\usepackage{amssymb,natbib}

\input{symbols}

\def\be{\begin{equation}}
\def\ee{\end{equation}}
\def\bea{\begin{eqnarray}}
\def\eea{\end{eqnarray}}

\begin{document}
\vspace*{4cm}
\title{MICROQUASARS AND JETS}

\author{ S. CHATY }

\address{AIM - Astrophysique Interactions Multi-\'echelles \\
(Unit\'e Mixte de Recherche 7158 CEA/CNRS/Universit\'e Paris 7 
Denis Diderot) \\
CEA Saclay, DSM/DAPNIA/Service d'Astrophysique, B\^at. 709 \\
L'Orme des Merisiers, FR-91 191 Gif-sur-Yvette Cedex, France
}

\maketitle\abstracts{
I present an overview of past, present and future research on
microquasars and jets, showing that microquasars, i.e. galactic
jet sources, are among the best laboratories for high energy
phenomena.  After remindind the analogy with quasars, I focus
on one of the best microquasar representatives, probably the
archetype, namely GRS 1915+105, and present accretion and ejection
phenomena, showing that only a multi-wavelength approach allows
a better understanding of phenomena occuring in these sources. Thereafter,
I review jets at different scales: compact jets, large-scale jets,
and the interactions between ejections and the surrounding medium. I
finish by speaking about microblazars and ultraluminous X-ray sources.
}

\section{Prelude to microquasars} \label{prelude}

In 1979 was discovered the microquasar prototype: SS 433, a
high-energy source exhibiting precessing jets at frame velocity $0.26
c$, with emission lines observed in the optical, showing that the jet
content was baryonic \citep{margon:1984}. SS 433 is surrounded by a
supernova remnant: W50, and there are clear signs of interaction
between SS 433 jets and W50 nebula (see e.g. \citeauthor{dubner:1998}
\citeyear{dubner:1998}).  The
question which arose was then: how can a galactic object eject matter
at such relativistic velocities ($\Gamma$=1.04)? This object exhibited
such unusual properties, that it was probably impossible to foresee
that, two decades later, jet sources would become quite common. SS 433
had everything of a microquasar, apart from the name.

\section{Youth of microquasars: analogy with quasars} \label{youth}

In 1990, the {\it SIGMA} telescope, orbiting on board {\it Granat}, was
launched. It was designed to observe galactic black hole candidates,
because its observing energy band corresponded to the energy released
by accretion around compact objects.  In 1992 the first so-called
microquasar, $\annihilateur$, was identified
\citep{mirabel:1992b}. This source was exhibiting bipolar radio jets
spread over several light-years. This was the first such
observation in our Galaxy, however jets had been already observed emanating
from distant galaxies. Therefore this observation made clear the
existence of a morphological analogy between quasars and microquasars.

\begin{figure}
\centerline{\psfig{file=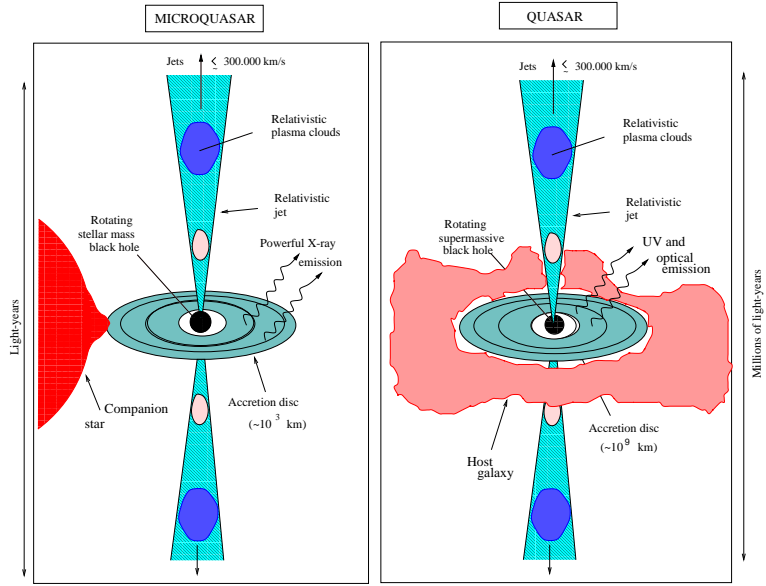,angle=-90.,width=10.cm}}
\caption[]{\label{quasar_microquasar} Schematic view illustrating 
analogies between quasars and microquasars \citep{chaty:1998}.  Note
the different mass and length scales between both types of objects.  }
\end{figure}

Although there is no clear definition of a microquasar, we can
characterize it as a galactic binary system --constituted of a compact
object (stellar mass black hole or neutron star) surrounded by an
accretion disc and a companion star-- emitting at high-energy and
exhibiting relativistic jets. A schematic view of a
microquasar, compared with quasars, is given in Figure
\ref{quasar_microquasar}. Taking this broad definition, we observed
nearly $20$ microquasars in our Galaxy, and it is one of the main
subjects of study by current space missions. Since each component
 of the system emits at different wavelengths, it is
necessary to undertake multi-wavelength observations in order to
understand phenomena taking place in these objects.

In 1992 the WATCH/GRANAT telescope discovered the black hole candidate
GRS 1915+105 \citep{castro:1994}, which would become the archetype of
microquasars.  Two years later, by observing this source with the VLA
(arcsec scale), \cite{mirabel:1994a} detected apparent superluminal
motions, while frame velocity was $v\sim0.92c$.  It became then
rapidly clear that the advantages of microquasars compared to quasars
were that i) they are closer, ii) it is possible to observe both
(approaching and receding) jets, and iii) the accretion/ejection timescale is
much shorter.  After this observation of superluminal motions, the
morphological analogy with quasars became stronger, and the question
was then, is this morphological analogy really sustended by physics? 
If the answer is yes, then microquasars really are
``micro''-quasars. For instance, there should exist microblazars
(microquasar whose jet points towards the observer), in order to
complete the analogy with quasars.

We will see in the following that this quasar/microquasar analogy
became rapidly very fruitful, the field of quasars benefitting of
microquasars, and vice versa. For instance, because accretion/ejection timescale is proportional to black hole mass, it is easier
(because faster) to observe accretion/ejection cycles in microquasars
than in quasars \footnote{Characteristic timescale of phenomena
occuring very close to the last stable orbit around the black hole of
mass $M$ is given by $\tau \sim \frac{\rg}{c} \sim M$, where $\rg$ is
the Schwarzschild radius.  Therefore, this timescale is proportional
to the mass of the black hole. If a stellar mass black hole exhibits
accretion/ejection cycles of a few minutes, a supermassive black hole
will exhibit corresponding cycles on a few thousands of years.}. On
the other hand the understanding of ejection phenomena in microquasars
have largely benefitted from jet models developped for active
galaxies.

\section{Maturity of microquasars: accretion and ejection}

\begin{figure}
\centerline{\psfig{file=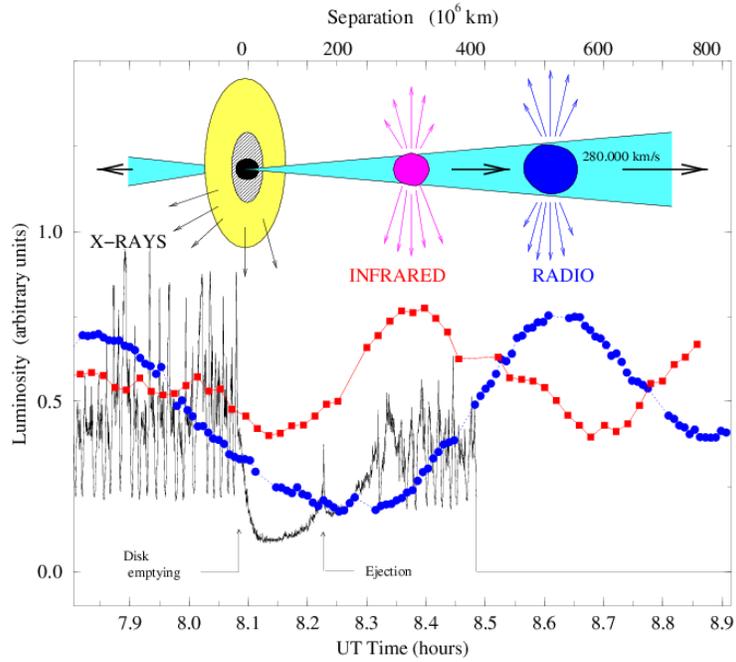,angle=-90.,width=10.cm}}
\caption[]{\label{grs_sept_french} Observation of the link between 
accretion and ejection. X-ray, NIR and radio lightcurves of $\grs$
during the 1997 September 9 multi-wavelength observation campaign
(\citeauthor{chaty:1998} \citeyear{chaty:1998}; 
\citeauthor{mirabel:1998a} \citeyear{mirabel:1998a}). The
disappearance of the internal part of the accretion disc (decrease in
the X-ray flux) is followed by an ejection of relativistic plasma
clouds (oscillation in the NIR and radio).}
\end{figure}

\begin{figure}
\centerline{\psfig{file=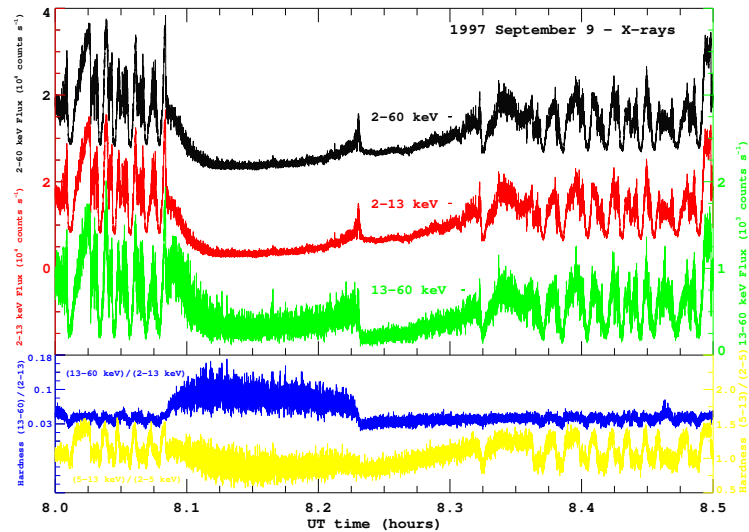,angle=90.,width=10.cm}}
\caption[]{\label{xte_9sept1997_zoom} Same observations as above, but only
X-ray observations are shown, and enlarged on the UT time interval
[8.0-8.5] hours. From top to bottom: X-ray flux in 2-60 keV, 2-13 keV
and 13-60 keV energy bands; hardness ratio $\frac{13 - 60 keV}{2 - 13
keV}$ and $\frac{5 - 13 keV} {2 - 5 keV}$ \citep{chaty:1998}.}
\end{figure}

GRS 1915+105 will once again play an important role in the
understanding of microquasars.  In 1997, after performing many
multi-wavelength observation campaigns of this source, the link
between accretion and ejection was discovered (\citeauthor{chaty:1998}
\citeyear{chaty:1998}; \citeauthor{mirabel:1998a} 
\citeyear{mirabel:1998a}). Examining Figure \ref{grs_sept_french}, 
we can see the disappearance of the internal part of the accretion
disc, shown by a decrease in the X-ray flux, followed by an ejection
of relativistic plasma clouds, corresponding to an oscillation in the
near-infrared (NIR) and then in the radio, the cloud becoming
progressively optically thin.  The analysis of X-ray fluxes and
hardness ratios, shown in Figure \ref{xte_9sept1997_zoom}, suggests
that it is mainly the part emitting at higher energy which is ejected
at the time of the X-ray spike. This supports the interpretation that
part of the corona (surrounding the compact object in the central part
of the accretion disc) is ejected during this cycle
\citep{chaty:1998}. Each of these accretion/ejection cycles last for 
$\sim10$ min, and they are recurrent, occuring every $\sim 30$--$45$
min.  Not only it is interesting to point out that these observations
had not been performed on quasars, even after nearly 40 years of
study, but also that for the first time microquasars were taking over
on the quasars, bringing new discoveries.  Five years later, similar
phenomena would be reported on the quasar 3C120, compiling 3 years of
observations \citep{marscher:2002}. These observations from both types
of objects confirmed that the morphological quasar/microquasar analogy
was sustended by physics \footnote{Another compelling evidence of this
analogy is given by the supermassive black hole at the center of our
Galaxy: with a mass of $3.6\times 10^{6}\Msol$ it exhibits a few tens
minutes NIR quasi-periodic oscillations (QPOs; \citeauthor{genzel:2003}
\citeyear{genzel:2003}),
when stellar mass black holes exhibit a few millisecond X-ray QPOs,
consistent with the mass ratio.}.

\section{The golden age of microquasars}

We do not discuss here the different accretion and ejection
models, but refer the reader to e.g. \cite{fender:2001a} for a
description of these models and how they relate to different ejection
states. We simply remind that the standard model is
constituted of thermal emission coming from a multicolour black body accretion
disc and of non-thermal emission of plasma corona, and that jets are
observed during low/hard states (historically referring to X-rays). 
Concurrent models invoke jet synchrotron emission from radio to
X-rays.  Therefore the main uncertainty in this domain concerns the
underlying physical process: comptonization or synchrotron? An
answer might be given by polarization observations.
High energy instruments do not allow this yet, 
and NIR polarimetric observations are still beginning.
\cite{dubus:2005} report NIR polarimetric
observations of the microquasar $\xtejqcq$, performed in 2003 at
ESO/NTT.  These observations were performed on the decline (at
$\sim2.5$ count/s) of a small amplitude outburst peak (4.5 count/s)
lasting about a month detected by {\it Rossi-XTE}/ASM \citep{sturner:2005}. In
NIR, it was 3.2 mag brighter than in quiescence. $\xtejqcq$
polarisation is inconsistent with other stars of the field of view at
the $2.5 \sigma$ level, suggesting an intrinsic NIR polarization
p=0.9--2.0\% perhaps due to synchrotron emission from the jet,
associated with the outburst
\citep{dubus:2005}.

\begin{figure}
\setlength{\unitlength}{1.0cm}
\begin{picture}(7.,7.)(-2.,-7.5)
\includegraphics{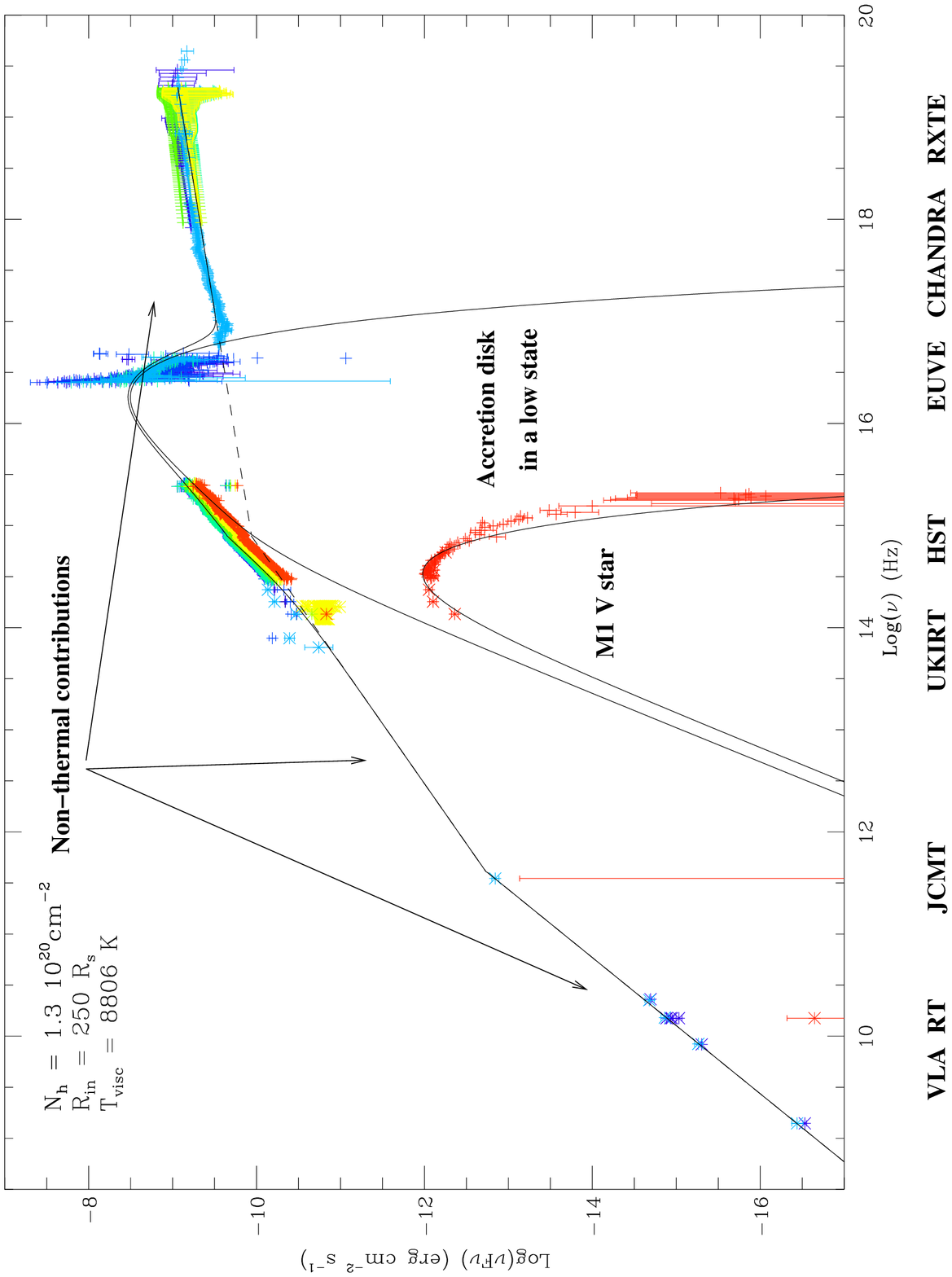}
\end{picture}
\caption[]{\label{sed_1118} Spectral energy distribution of the 
microquasar $\xtejodh$ \citep{chaty:2003b}.}
\end{figure}

To understand accretion/ejection models, it is therefore necessary to
undertake a multiwavelength approach and get the spectral energy
distribution (SED) of various sources.  There is a small number of
microquasars for which this has been done intensively, the jet source
and black hole XTE J1118+480 being one of them, favored by its
very low absorption on its line of sight \citep{chaty:2003b}. In Figure
\ref{sed_1118} I report the SED of this
source, including 6 different epochs of simultaneous multi-wavelength
observations from radio to X-rays, performed with 8
different instruments. On this Figure I overplot the thermal
emission of the multicolour black body accretion disc, the 
emission from the companion star, and non-thermal emission which
appears to be necessary to account for radio, NIR and X-ray
domains.  In \cite{chaty:2003b} it has been shown, by using a
non-linear Monte-Carlo simulation, that the presence of hot spherical
plasma in the centre can account for the emission of the source from
optical to X-rays.  However other models show that this emission can
also be described by a jet emitting from radio to X-rays, as in the
case of active galaxies \citep{markoff:2001}.  This question about the
jet contribution is therefore still a matter in the debate.

It is interesting to compare XTE J1118+480 and GRS 1915+105
SEDs. During large multiwavelength campaigns from radio to hard
X-rays, \cite{ueda:2002} and
\cite{fuchs:2003} have shown the presence of a flat radio spectrum,
during the ``plateau'' (or low/hard) state of GRS 1915+105.  They
also confirm that the jet contributes to the emission in the
NIR domain. A comparison of the accretion/outflow energy ratio of
both sources XTE J1118+480 \& GRS 1915+105 shows that they both fall
into the regime of radio-quiet quasars \citep{chaty:2003b}.

Simultaneous multi-wavelength observations of both types of objects,
namely microquasars and quasars, will eventually bring severe
constraints on accretion-ejection models (e.g. Blandford-Payne,
Blandford-Znajek, Magneto-Rotational Instability...), and on the
nature of the jets (are they made of $e^-/e^+$ or $e^-/p$?). For instance,
putting together radio and X-ray observations suggests that a coupling
exists between both domains, $F_{\rm rad} \propto F_{\rm X}^{+0.7}$, for
galactic \citep{gallo:2003} and extragalactic jet sources
\citep{falcke:2004}, but a
good understanding of this coupling still misses.
Some answers might also come from the detection of (Doppler- shifted?) 
annihilation emission lines, and also from observations of QPOs in
microquasars.

\section{The hidden face of microquasars: jets and surroundings...}

Jets of microquasars can be observed at different scales,
corresponding to different sizes and energy outputs involved.
Observations of sporadic ejections at large scale were performed
first, as described in Section \ref{youth}.  A steady compact jet
has been observed in a few microquasars, for instance in GRS 1915+105
(at the milli-arcsec scale, where 10mas = 1AU;
\citeauthor{dhawan:2000b} \citeyear{dhawan:2000b};
\citeauthor{fuchs:2003} \citeyear{fuchs:2003};
\citeauthor {ribo:2004} \citeyear{ribo:2004}).
Since these jet sources eject a large amount of matter in the
interstellar space, which is far from being empty, it appears fruitful
to look for interactions between jets and surroundings of the
microquasar.  The first example is $\annihilateur$, which exhibits a
steady jet, probably due to the braking of its continuous jet in the
interstellar medium.  The signature of such an interaction might be
the observation, directly in the jets, of a narrow annihilation line
at 511 keV, due to $e^+$ colliding with the interstellar medium
\footnote{Annihilation lines have been reported on this source
(therefore also called ``the great annihilator of the Galaxy'') but
likely coming from the central source \citep{bouchet:1991}.}.
Large-scale jets are now regularly observed in X-rays. \cite{corbel:2002} have
observed such jets emanating from the microquasar $\xtejqcq$, at 
$45\arcsec$ of the central source. To emit at such energy,
the particles have to be accelerated up to TeV energy, again
strengthening the analogy with quasars.  

By studying the interactions between jets and the interstellar medium, one
can not forget GRS 1915+105: always active, transient, and the place
of very energetic ejections.  Such interactions in the surroundings of
GRS 1915+105 had already been suggested nearly 10 years ago by
\cite{mirabel:1996a}.  In August 1995, during a strong and long X-ray
outburst of GRS 1915+105, the radio source was resolved in 2 jets, and
the NIR emission increased significantly between 2 and 5 days after
the radio burst. \cite{mirabel:1996a} interpreted this as the presence
of an extended cocoon of dust, heated by ejections. However it was
unknown if the cocoon had been created by previous ejections, or by
accumulation of ISM dust.  This dust in the surroundings of this
microquasar was later confirmed by {\it Chandra} \citep{lee:2002} and
{\it ISO} \citep{fuchs:2001} observations.  And what about the
surroundings of GRS 1915+105, at larger scale?  A low-resolution centimeter
map exhibits two sources aligned with the central source
\citep{chaty:2001a}. By observing them at higher resolution, it
appeared a strange non-thermal feature in the south-east lobe, which
might be a synchrotron signature of interactions between jets and
ISM. However, \cite{chaty:2001a} concluded that even if, based on the
energy output, the interaction is a possibility, there is no
observational fact allowing to confirm that this strange feature is
the signature of interaction between jets and interstellar medium.

Finally, all these observations of jets bring us to another important
question in the field of microquasars: are the jets a propagation
of plasma clouds or the propagation of shock waves?  The first
interpretation is usual among the microquasar community, and the
second one among the extragalactic community. By applying 3C273 model
to GRS 1915+105, \cite{turler:2004} have shown that ejections in GRS
1915+105 could be described as the propagation of a shock wave forming
at 1AU, with dissipative stream at $v=0.6c$.

\section{Ubiquity of microquasars}

Even if microquasars are not everywhere, they are more and more
present!  We have seen in Section \ref{youth} that microblazars should
exist if the analogy with quasars was sustended by physics. However
the problem with microblazars is that they are difficult to observe
since flares, although strong, are short \footnote{For a microblazar
with a jet frame velocity $v=0.98c$, the time of the outburst is
shortened by a factor 10, the flux multiplied by 1000, and the photon
energy increased compared to a microquasar.}.  Jet precession could
produce intermittent microblazars (see e.g. \citeauthor{massi:2004}
\citeyear{massi:2004}).  There are some hints that some microblazars
have been observed.  The source V4641 Sgr exhibited a one-day flare,
becoming for a short time the brightest X-ray source of the sky,
increasing from 1.6 to 12.2 Crab, and in optical from $14$ to $8.8$
magnitudes, exhibiting a wind velocity of 5000 km/s
\citep{chaty:2003a}.  This source was claimed to be a microblazar,
since at a distance of 6 kpc, the jets would have had an apparent
velocity of $v\sim10c$.  However this apparent velocity is based on
the uncertain motion of the radio lobe, due to lack of good observing
coverage, therefore there is no conclusive evidence that this source
is a microblazar.

Ultra-luminous X-ray sources (ULXs) are observed near active galactic nuclei 
at high stellar formation rates. Are they beamed jets from microquasars
(and in this case microblazars would not be missing to the family
anymore),
or black holes of intermediate mass ($\sim1000 M_{sol}$) \citep{kording:2002}?
Are there ULXs in our Galaxy? There are claimed associations between 
galactic microquasars and gamma-ray sources: e.g. LS 5039 \citep{paredes:2000},
and hundreds of unidentified gamma-ray sources still exist...
so this field is still full of discoveries to come.
 
\section*{Acknowledgments}
I acknowledge Elmar K\"ording, Leonardo Pelliza and Marc Rib\'o for 
a careful rereading of the manuscript and useful discussions.


\end{document}

%% file: symbols.tex
\def\annihilateur{1 \mbox{E 1740.7-2942}}

\def\grs{\mbox{GRS 1915+105}}

\def\xtejodh{\mbox{XTE J1118+480}}
\def\xtejqcq{\mbox{XTE J1550-564}}

\def\Msol{\mbox{ }M_{\odot}}

\def\arcsec{^{\prime \prime}}

\def\rg{r_{\scriptsize g}}
\def\ltsima{\; \buildrel < \over \sim \;}
\def\simlt{\lower.5ex\hbox{\ltsima}}            
\def\gtsima{\; \buildrel > \over \sim \;}
\def\simgt{\lower.5ex\hbox{\gtsima}}            
